\documentstyle[12pt,epsfig]{article}

\newlength{\dinwidth}                                                    
\newlength{\dinmargin}
\def\lapproxeq{\lower .7ex\hbox{$\;\stackrel{\textstyle                                                    
<}{\sim}\;$}}                                                    
\def\gapproxeq{\lower .7ex\hbox{$\;\stackrel{\textstyle                                                    
>}{\sim}\;$}}                                                    
\def\be{\begin{equation}}                                                    
\def\ee{\end{equation}}                                                    
\def\bea{\begin{eqnarray}}                                                    
\def\eea{\end{eqnarray}}

\begin{document}                                                    
\titlepage                                                    
\begin{flushright}                                                    
IPPP/08/93   \\
DCPT/08/186 \\                                                    
\today \\                                                    
\end{flushright}                                                    
                                                    
\vspace*{2cm}                                                    
                                                    
\begin{center}                                                    
{\Large \bf Soft processes at the LHC}\\   
\vspace*{0.5cm}
{\Large \bf II: Soft-hard factorization breaking and gap survival}                                                    
                                                    
\vspace*{1cm}                                                    
M.G. Ryskin$^{a,b}$, A.D. Martin$^a$ and V.A. Khoze$^{a,b}$ \\                                                    
                                                   
\vspace*{0.5cm}                                                    
$^a$ Institute for Particle Physics Phenomenology, University of Durham, Durham, DH1 3LE \\                                                   
$^b$ Petersburg Nuclear Physics Institute, Gatchina, St.~Petersburg, 188300, Russia            
\end{center}                                                    
                                                    
\vspace*{2cm}                                                    
                                                    
\begin{abstract}                                                    
We calculate the probability that the rapidity gaps in diffractive processes survive both eikonal and enhanced rescattering. We present arguments that enhanced rescattering, which violates soft-hard factorization, is not very strong. Accounting for NLO effects, there is no reason to expect that the black disc regime is reached at the LHC. We discuss the predictions for the survival of the rapidity gaps for exclusive Higgs production at the LHC. 

\end{abstract}    

\section{Overview of rapidity gap survival}

\begin{figure}
\begin{center}
\includegraphics[height=4cm]{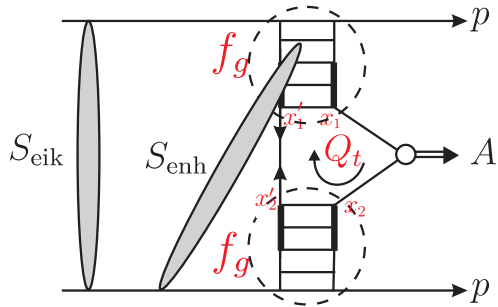}
\caption{The mechanism for the exclusive process $pp \to p+A+p$, with the eikonal and enhanced survival factors shown symbolically. The thick lines on the Pomeron ladders, either side of the subprocess ($gg \to A$), indicate the rapidity interval $\Delta y$ where enhanced absorption is not permitted \cite{thr,JHEP,kkmr}.}
\label{fig:pAp}
\end{center}
\end{figure}

We discuss the effect of screening corrections (that is, the rapidity gap survival factors) in diffractive processes. In particular we study the exclusive central production of Higgs bosons or another small-size heavy system, which we collectively denote by $A$. The process we have in mind is $pp \to p+A+p$, where the $+$ signs indicate rapidity gaps. Usually the cross sections of such processes are calculated assuming soft-hard factorization. Symbolically it can be written as
\be
\sigma~=~|f_g \otimes {\cal M}_A \otimes f_g |^2 \cdot S^2,
\label{eq:sig}
\ee
 where the first factor is the bare amplitude for the exclusive process calculated in perturbative QCD as the convolution of the hard matrix element ${\cal M}_A$ and the gluon distributions $f_g$. This is shown in the right half of Fig.~\ref{fig:pAp}. The factor $S^2$ is the probability that the secondaries, which are produced by soft rescattering, do not populate the gap. 

As mentioned above, the cross section in form (\ref{eq:sig}) assumes soft-hard factorization. The survival factor, calculated from a model of soft interactions, does not depend on the structure of the original perturbative QCD amplitude; symbolically it is shown by $S_{\rm eik}$ in the left part of the figure.

Actually, the situation is more complicated. We need to consider so-called ``enhanced'' rescattering corrections which involve intermediate partons, as indicated by the typical contribution to $S^2_{\rm enh}$ shown in Fig.~\ref{fig:pAp}. As a rule, the partons which participate in the hard subprocess have large virtuality in transverse momentum and the probability of rescattering of such partons is negligible. Due to the strong $k_t$-ordering, the most important rescattering involves partons near the beginning of the evolution, that is near the input scale $q_0$. However, the value of $q_0$ is not fixed. To be able to neglect all the enhanced absorptive corrections we have to choose large $q_0$. On the other hand, the development of the parton shower starts from a low scale $q'_0 \sim 1/R$, where $R$ is the proton radius. This part of the evolution of the parton shower from the beam may be affected by the interaction with the target (and vice versa). In this way `enhanced' rescattering violates soft-hard factorization, leading to a distortion of the input distribution at $q_0$. Our aim is to quantify the rescattering effect which originates in the interval $(q'_0,~q_0)$. To do this we need to introduce components, $a$, of the Pomeron of different size. Here, we take three components, $a=P_1,~P_2,~P_3$ corresponding to large-, intermediate- and small-size as in the preceding paper \cite{KMRnns}. These were associated with transverse momenta of about 0.5, 1.5 and 5 GeV respectively.

\begin{figure}
\begin{center}
\includegraphics[height=3cm]{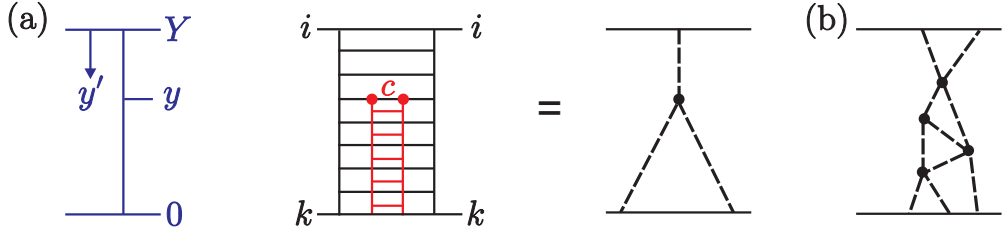}
\caption{(a) The ladder structure of the triple-Pomeron amplitude between diffractive eigenstates $|\phi_i\rangle,|\phi_k\rangle$ of the proton; the rapidity $y$ spans an interval 0 to $Y={\rm ln}s$. (b) A multi-Pomeron diagram.} 
\label{fig:2lad}
\end{center}
\end{figure}

The opacity of the `target $k$' (or beam $i$) proton with respect to intermediate parton $c$, generated by Pomeron component $a$ is denoted by $\lambda\Omega_k^a$ (or $\lambda\Omega_i^a$). The parameter $\lambda$ relates the triple-Pomeron coupling to the Pomeron coupling to the proton: $g_{PPP} \equiv \lambda g_N$. Its presence will allow multi-Pomeron-Pomeron absorptive effects to be included \cite{KMRnns}. If, for the moment, we consider a Pomeron with only one component, then the opacity satisfies the evolution equation in rapidity $y$, at impact parameter $b$,
\begin{equation}
\frac{d\Omega_k(y,b)}{dy}\,=\,e^{-\lambda\Omega_i(y',b)/2}~~~e^{-\lambda
\Omega_k(y,b)/2}~~
\left(\Delta+\alpha'\frac{d^2}{d^2b}\right)\Omega_k(y,b)\; ,
\label{eq:evol1}
\end{equation}
where $y'={\rm ln}s-y$. Let us explain the meanings of the three factors on the right-hand-side of (\ref{eq:evol1}). If only the last factor, (...)$\Omega_k$, is present then the evolution generates the ladder-type structure of the bare Pomeron exchange amplitude, where the Pomeron trajectory $\alpha_P=1+\Delta+\alpha't$. The inclusion of the preceding factor allows for rescatterings of an intermediate parton $c$ with the ``target'' proton $k$; Fig.~\ref{fig:2lad}(a) shows the simplest (single) rescattering which generates the triple-Pomeron diagram. Finally, the first factor allows for rescatterings with the beam $i$. In this way the absorptive effects generated by all multi-Pomeron-Pomeron diagrams are included, like the one shown in Fig.~\ref{fig:2lad}(b) containing, in general, $m \to n$ multi-Pomeron vertices. 
There is an analogous equation for the evolution of $\Omega_i(y',b)$, and the two equations may be solved iteratively.  

Now, consider a Pomeron with more components $a$, then these equations have a matrix form in $aa'$ space \cite{KMRnns}. At each step of the evolution the equation includes absorptive factors of the form $e^{-\lambda\Omega^a(y,b)/2}$. By solving the equations with and without these suppression factors, we can quantify the effect of enhanced absorption. Moreover, we start the evolution from the large component $P_1$, and since the evolution equations allow a transition from one component to another (corresponding to BFKL diffusion \cite{BFKL} in ln$k_t$ space), we can see how the enhanced absorption will affect the distribution of high $k_t$ in the small size component $P_3$.

\begin{figure}
\begin{center}
\includegraphics[height=1.5cm]{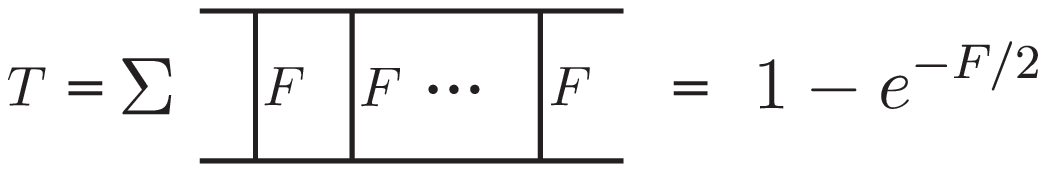}
\caption{Pure elastic eikonal scattering.} 
\label{fig:T}
\end{center}
\end{figure}
\begin{figure}
\begin{center}
\includegraphics[height=4cm]{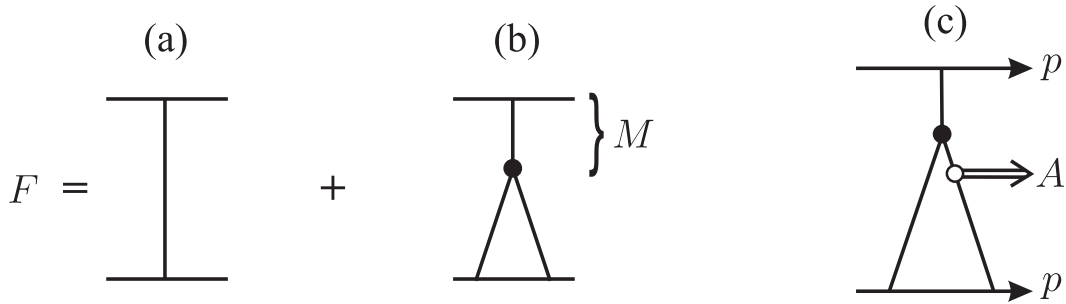}
\caption{(a) The bare pole amplitude; (b) the triple-Pomeron diagram describing high mass, $M$, diffractive dissociation; (c) the contribution of the triple-Pomeron coupling to central exclusive production.} 
\label{fig:F}
\end{center}
\end{figure}
If we consider pure elastic eikonal rescattering, Fig.~\ref{fig:T}, then phenomenologically we can describe the bare high-energy elastic amplitude $F$ by just one Regge pole, see Fig.~\ref{fig:F}(a). On the other hand, we know that a sizeable part of the cross section comes from diffractive dissociation.  High-mass, $M$, diffractive dissociation is usually described by triple-Pomeron diagram, and we have to include this contribution both in the irreducible amplitude $F$ (that is, the $pp$ opacity) as in Fig.~\ref{fig:F}(b), and in the bare amplitude for central exclusive production, as in Fig.~\ref{fig:F}(c).  The latter corresponds to a contribution to the so-called enhanced screening factor $S_{\rm enh}$ of Fig~\ref{fig:pAp}. Physically, this factor accounts for the absorption of intermediate partons inside the original central exclusive amplitude. It is called {\it enhanced} since we have to multiply the probabilities of absorption on each individual intermediate parton, and thus the final effect is {\it enhanced} by the large multiplicity of intermediate partons.

A complication is that $S_{\rm enh}$ cannot be considered simply as an overall multiplicative factor. The probability of interaction with a given intermediate parton depends on its position in configuration space; that is, on its impact parameter ${\mathbf b}$ and its momentum $k_t$. This means that $S_{\rm enh}$ simultaneously changes the distribution of the active partons which finally participate in the hard subprocess.

We are now in a position to calculate the rapidity gap survival factors including not just {\it eikonal} rescattering, but also allowing for the possibility of {\it enhanced} screening.

\section{Eikonal screening}

The gap survival factor caused by {\it eikonal} rescattering of the
Good-Walker eigenstates \cite{GW}, for a fixed impact paramter ${\mathbf b}$, is
\begin{equation}  S^2_{\rm eik}({\mathbf b})~ = ~\frac{\left| {\displaystyle\sum_{i,k}} |a_{i}|^2~|a_{k}|^2~{\mathcal
M}_{ik}({\mathbf b})~\exp(-F_{ik}(s,{\mathbf b})/2)\right|^2}{\left|{\displaystyle \sum_{i,k}}
|a_{i}|^2~|a_{k}|^2~{\mathcal M}_{ik}({\mathbf b}) \right |^2} \,.
\label{eq:c3pp}
\end{equation}
The $F_{ik}$ are the $s$-channel two-particle {\it irreducible} amplitudes for the scattering of the various diffractive eigenstates\footnote{Recall that, in the global model of soft interactions of the preceding paper \cite{KMRnns}, the Good-Walker formalism \cite{GW} was used to decompose the proton wave function  in terms of three diffractive eigenstates $|p\rangle = \sum_i a_i |\phi_i \rangle$. That is, we used a three-channel eikonal to describe elastic rescattering and low-mass proton dissociation.} $|\phi_i\rangle$ and $\phi_k\rangle$, for given separations  ${\mathbf b}={\mathbf b}_1-{\mathbf b}_2$
 between the incoming protons
  \begin{equation}
 F_{ik}(Y,{\mathbf b})=\frac
 1{\beta_0^2}\sum_a\int\Omega^a_{k}(y,{\mathbf b}_1,{\mathbf b}_2)\Omega^a_{i}(Y-y,{\mathbf b}_1,{\mathbf b}_2)
 d^2b_1d^2b_2\delta^{(2)}({\mathbf b}_1-{\mathbf b}_2-{\mathbf b})
 \label{eq:F}
 \end{equation}
 where $Y=\ln s$. It is the total opacity of the $ik$ interaction. It includes the sum over the different Pomeron components $a$ and the integral over the impact parameter which accounts for the overlap of the partonic wave functions (that is, the opacities $\Omega_i$ and $\Omega_k$) generated by the beam and target protons for fixed ${\mathbf b}={\mathbf b}_1-{\mathbf b}_2$ separation. The factor $1/\beta_0^2$ provides the appropriate normalisation as explained in the preceding paper \cite{KMRnns}.

The most topical case is Central Exclusive Diffractive (CED) production of a
Higgs boson (or other heavy system of small size) \cite{KMRpr,KMR,oldsoft,DKMOR}. Then the
matrix element ${\cal M}_{ik}$ of the hard subprocess $gg \to A$ can be calculated perturbatively, since
the large mass and small size of the centrally produced object introduces a
large scale \cite{KMRpr,KMR}. However, the shape of ${\cal M}_{ik}$ in
impact parameter space ${\mathbf b}$ and the relative couplings of the hard
matrix element to different Good-Walker eigenstates $|\phi_i\rangle$ are
not known. Their calculation is beyond the ability of perturbative QCD. 
 
One possibility is to say that the ${\mathbf b}$ dependence of ${\cal M}$
should be, more or less, the same as for diffractive
$J/\psi$ electroproduction ($\gamma+p\to J/\psi+p$), and the coupling
to the $|\phi_i \rangle$ component of the proton should be proportional to the same factor 
$\gamma_i$, as in a soft interaction. This leads to
\begin{equation}
{\cal M}_{ik}\propto \gamma_i\gamma_k\exp(-b^2/4B)
\label{eq:m1}
\end{equation}
with $B\simeq 4$ GeV$^{-2}$ \cite{JPsi}.
Under these assumptions, the `soft' model of \cite{KMRnns} predicts 
$S^2_{\rm eik}$=0.017 for the LHC energy, which is close to the value 0.023 obtained recently by the Tel-Aviv
group \cite{GLMM}. 

Expressing the survival factors in this manner is too simplistic. However, these numbers are frequently used as the reference point. This can be misleading, since $S^2$ calculated in this way strongly depends on an additional assumption about the slope $B$, a quantity which is mainly of `soft' origin.

\begin{figure}
\begin{center}
\includegraphics[height=7cm]{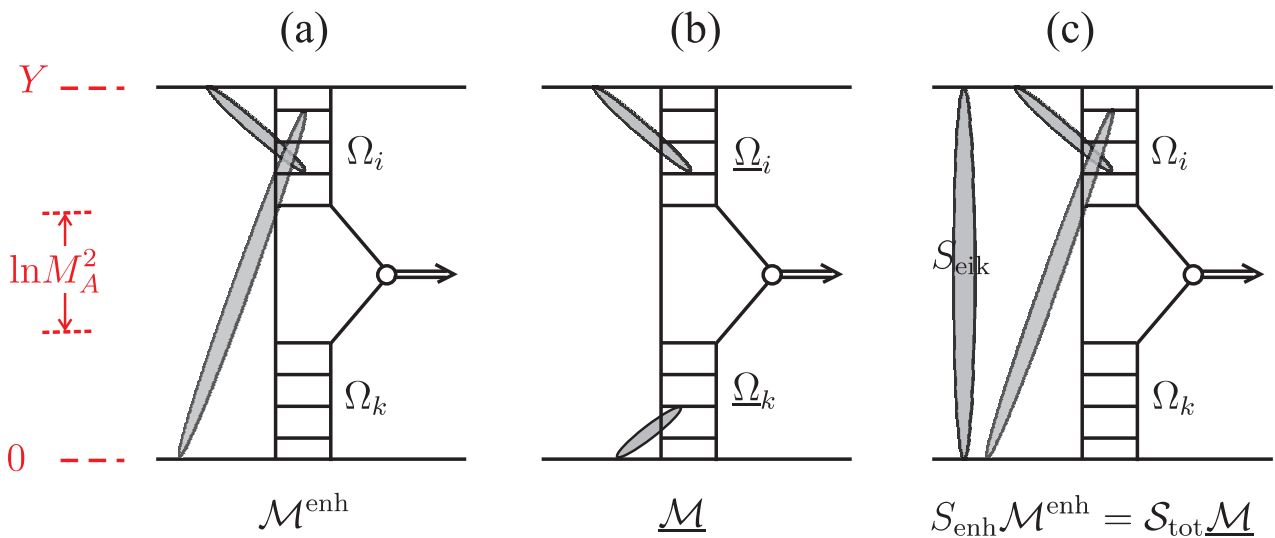}
\caption[*]{(a): A symbolic representation of (\ref{eq:M}); (b) and (c): diagrams relevant for the computation of $S^2_{\rm enh}$ and $S^2_{\rm eik}S^2_{\rm enh}$ of (\ref{eq:sym1}) and (\ref{eq:sym2}), where $\underline\Omega$ is the opacity when the screening corrections embracing the hard matrix element are neglected.}
\label{fig:SS}
\end{center}
\end{figure}

Another, more plausible and self-consistent, possibility is to calculate ${\cal M}_{ik}({\mathbf b})$ under the assumption that the active gluons that participate in the hard subprocess come from the small-size component $P_3$ of the Pomeron. That is,
\begin{equation}
 {\cal M}_{ik}^{\rm enh}({\mathbf b})=\frac{M(gg\to A)}
 {\beta_0^2}\int\Omega^{P_3}_i(y_1,{\mathbf b}_1,{\mathbf b}_2)\Omega^{P_3}_k(y_2,{\mathbf b}_1,{\mathbf b}_2)
 d^2b_1d^2b_2\delta^{(2)}({\mathbf b}_1-{\mathbf b}_2-{\mathbf b})\; ,
 \label{eq:M}
 \end{equation}
where the rapidities $y_j=-{\rm ln}\xi_j$ with $j=1,2$, correspond to the momentum fractions carried by the active gluons which participate in the hard subprocess with matrix element $M(gg\to A)$. The superscript ``enh'' is to indicate that the matrix element calculated in this way already accounts for the ``enhanced'' rescattering on intermediate partons.
A symbolic representation of the formula is given in Fig.~\ref{fig:SS}(a). The formula is similar to formula (\ref{eq:F}) for the irreducible amplitude $F_{ik}({\mathbf b})$. The difference is that now the two active gluons are separated by a large rapidity interval ln$M_A^2$, so that $y_1+y_2=Y-{\rm ln}M_A^2$.

\section{Enhanced screening}
However, as mentioned above, besides pure eikonal screening, there may be {\it enhanced} rescattering arising from absorption
of the intermediate partons in the hard matrix element
${\cal M}_{ik}$. Working at LO (of collinear approximation) we have to
neglect such an effect. Due to strong $k_t$-ordering the
transverse momenta of all the intermediate partons are very large
(i.e. the transverse size of the Pomeron is very small) and therefore the absorptive
effects are negligible. Nevertheless, this may be not true at a very low
$x$, where the parton densities become close to saturation and the
small value of the absorptive cross section is compensated by the large
value of the parton density.

Indeed, the contribution of the first {\it enhanced} diagram, which
describes the absorption of an intermediate parton, was estimated in
the framework of the perturbative QCD in Ref.\cite{BBKM}. It turns out
that it could be quite large. On the other hand, such an effect is not seen
experimentally. The absorptive corrections due to enhanced
screening must increase with energy. This is not observed in the
present data (see \cite{JHEP} for a more detail discussion).
One reason is that the gap survival factor $S^2_{\rm eik}$
already absorbs almost all the contribution
from the centre of the disk. The parton essentially only survives eikonal
rescattering on the periphery, that is at large impact parameters $b$. On the other hand,
on the periphery, the parton density is rather small and the probability
of {\it enhanced} absorption is not large. This fact can be seen in
Ref. \cite{Watt}. There the momentum, $Q_s$, below which we may approach
saturation, was extracted from the HERA data in the framework of the
dipole model. Already at $b=0.6$ fm the value of $Q^2_s<0.3$ GeV$^2$
for the relevant $x\sim 10^{-6}$.

Recall that, in our model, the absorption caused by rescattering of intermediate partons is included as exponential factors, $e^{-\lambda(\Omega_i+\Omega_k)/2}$, in equations (\ref{eq:evol1}) for the opacities. So we have the possibility to quantify the role of this (`enhanced') exponential suppression by solving the evolution equations with and without the exponential factors. 

First, we could calculate the value of ${\cal M}_{ik}$ {\it excluding}
 any effect of `enhanced' absorption. To do this we would delete the factors
 $e^{-\lambda(\Omega_k+
\Omega_i)/2}$ from the equation (\ref{eq:evol1}), and then
 solve the modified equations for opacities $\Omega_i$ and $\Omega_k$
 respectively. However, this is not what we need, and before we start the calculation it is necessary to note two pragmatic observations.  The problem is that we no longer have exact factorisation between the hard and soft parts of the process. Thus before computing the effect of soft absorption we must fix what is included in the bare CED amplitude calculated in terms of perturbative QCD.

\section{Matching $S^2$ with the hard matrix element}

The first observation is that the bare amplitude is calculated as a convolution of two generalised (skewed) gluon distributions with the hard subprocess matrix element \cite{KMRpr,KMR}. These gluon distributions are determined from integrated gluon distributions of a global parton analysis of mainly deep inelastic scattering data. However, these phenomenological integrated parton distributions already include the interactions of the intermediate partons with the parent proton. As a consequence, in order to determine $S^2_{\rm enh}$ we should compare the matrix element calculated including all enhanced diagrams, like the one sketched in Fig.~\ref{fig:SS}(a), with the matrix element obtained by excluding, in the evolution of $\Omega_i$, the exponential factor $e^{-\lambda\Omega_k/2}$ which `embraces' the hard subprocess, and keep only $e^{-\lambda\Omega_i/2}$, as indicated in Fig.~\ref{fig:SS}(b); and vice versa for the evolution of $\Omega_k$. In this way we mimic the gluons used in the perturbative QCD calculation of the cross section.  Let us denote the opacities, obtained this way, as
 $\underline{\Omega}$ and the corresponding matrix element of (\ref{eq:M}) as
 $\underline{\cal M}_{ik}({\mathbf b})$. These opacities mimic the phenomenological gluon distributions used in the perturbative QCD calculation. So, for fixed indices $i,k$, the enhanced survival factor is given by
\be
S^2_{\rm enh}~=~\frac{\left| {~\cal M}^{\rm enh} \right|^2}{\left| ~\underline{\cal M}~ \right|^2}.
\ee 

Now we are ready to calculate the survival factors. We would like to compare the survival factors computed as indicated in Fig.~\ref{fig:SS}(b) and (c). First, corresponding to Fig.~\ref{fig:SS}(c), we compute $S^2_{\rm eik}({\mathbf b})$ from (\ref{eq:c3pp}) where ${\cal M}_{ik}^{\rm enh}({\mathbf b})$ includes the full enhanced screening, but not the eikonal rescattering. Symbolically we have
\be
S^2_{\rm eik}({\mathbf b})~=~\frac{\left| {~\cal M}^{\rm enh}~*~e^{-F_{ik}/2}~ \right|^2}{\left| {~\cal M}^{\rm enh}~ \right|^2},
\label{eq:sym1}
\ee
where ${\cal M}^{\rm enh}$ is given by (\ref{eq:M}). Then we compute the {\it total} absorptive effect $S^2_{\rm enh}({\mathbf b})~S^2_{\rm eik}({\mathbf b})$ by replacing ${\cal M}^{\rm enh}$ in the denominator by $\underline{\cal M}$. The calculation of $\underline{\cal M}$ corresponds to Fig.~\ref{fig:SS}(b). That is,
\be
S^2_{\rm tot}({\mathbf b})~\equiv~S^2_{\rm enh}({\mathbf b})~S^2_{\rm eik}({\mathbf b})~=~\frac{\left| ~{\cal M}^{\rm enh}~*~e^{-F_{ik}/2}~ \right|^2}{\left| ~\underline{\cal M}~ \right|^2}.
\label{eq:sym2}
\ee

The second observation is that the phenomenologically determined generalised gluon distributions are usually taken at $p_t=0$, and then the observed ``total'' cross section is calculated by integrating over $p_t$ of the recoil protons {\it assuming} an exponential behaviour $e^{-Bp_t^2}$; that is
\be
\sigma ~=~\int\frac{d\sigma}{dp_{1t}^2 dp_{2t}^2}dp_{1t}^2 dp_{2t}^2 ~=~\frac{1}{B^2}\left.\frac{d\sigma}{dp_{1t}^2 dp_{2t}^2}\right|_{p_{1t}=p_{2t}=0}~,
\ee
where
\be
\int dp^2_t~e^{-Bp_t^2}~=~1/B~=~\langle p_t^2\rangle.
\ee
However, the total soft absorptive effect changes the $p_t$ distribution in comparison to that for the bare cross section determined from perturbative QCD. Moreover, the correct $p_t$ dependence of matrix element (\ref{eq:M}) does {\it not} have an exponential form. Thus the additional factor introduced by the soft interactions is not just the gap survival $S^2$, but rather $S^2\langle p^2_t \rangle^2$, where the square arises since we have to integrate over the $p_t$ distributions of {\it two} outgoing protons. Indeed in all the previous calculations the soft prefactor had the form\footnote{At larger impact parameter $b$ the absorption is weaker. Hence the value of $S^2$ increases with the slope $B$. It was shown that the ratio $S^2/B^2$ is approximately stable for reasonable variations of $B$ \cite{KMRSchi}.} $S^2/B^2$.  Now, in our model the ${\mathbf b}$ behaviour of ${\cal M}({\mathbf b})$ is driven by the opacities. Thus we present the final result in the form $S^2\langle p^2_t \rangle^2$, and so to compare it with previous predictions obtained for $B=4~{\rm GeV}^{-2}$ we need to introduce the ``renormalisation'' factor $(\langle p_t^2 \rangle B)^2$. The resulting (effective) value is denoted by $S^2_{\rm eff}$.

\section{Impact parameter dependence of $S^2_{\rm enh}$}
\begin{figure} [t]
\begin{center}
\includegraphics[height=8cm]{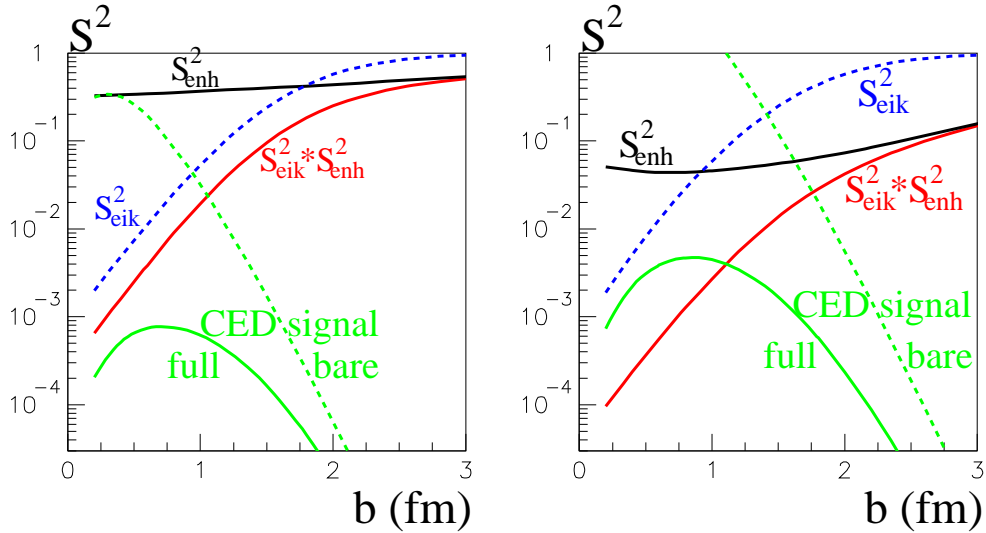}
\caption[*]{The survival factors for the Central Exclusive Diffractive (CED) production of a heavy $0^+$ system of mass $M_A$=150 GeV and $M_A$= 10 GeV.}
\label{fig:SSb}
\end{center}
\end{figure}

First, we present the results of the calculation of the survival factors at fixed $pp$ impact parameter ${\mathbf b}={\mathbf b}_1-{\mathbf b}_2$; that is, before integrating over $d^2b$. The results are presented in Fig.~\ref{fig:SSb} for the LHC energy $\sqrt{s}=14$ TeV for, first, the production of a Higgs boson of mass $M_A=120$ GeV and, second, for the central producton of an object of mass $M_A=10$ GeV (which may be a pair of high-$E_T$ photons or a $\chi_b$). For the 10 GeV object we assume that the active gluons come from the intermediate Pomeron state, $P_2$. As expected the eikonal suppression, $S^2_{\rm eik}$, is almost the same for both objects (with a bit stronger suppression for $P_3$ which lies in a bit smaller ${\mathbf b}$ domain). On the other hand, the enhanced suppression, $S^2_{\rm enh}$, is stronger for the lighter object as the rapidity interval allowed for the interaction with the intermediate partons is considerably larger, and the absorption cross section is larger for a lower $k_t$ parton.

After we multiply the bare QCD amplitude, shown by the dashed line in the plots of Fig.~\ref{fig:SSb}, by the total survival factors we obtain the impact parameter profiles of the final CED signal which peak around 0.8 fm. The signal vanishes as $b \to 0$ simply because we plot the integrand $b|{\cal M}|^2$ of $\int d^2b~S^2|{\cal M}({\mathbf b})|^2=2\pi \int b|{\cal M}({\mathbf b})|^2 db$; the normalisation of the signal in the figure was chosen to facilitate the plot.

Assuming that all the partons are distributed homogeneously, we would expect $S^2_{\rm enh} \to 1$ at the large values of $b$ of the periphery. However, the results show only a weak tendency to increase at large $b$. Thus, in the `soft' model of \cite{KMRnns} (with a small value of $\alpha'_P$), $S^2_{\rm enh}$ comes mainly from ``hot-spots'' in which many individual intermediate partons are concentrated within small $b$ domains. In other words, most of enhanced absorption occurs within the same parton shower, and is due to secondaries produced during the evolution with the same impact parameter $b$. On the other hand, a detailed analysis \cite{Watt} of HERA data shows that the value of the saturation scale $Q_s(b)$ decreases rapidly with increasing $b$. ($Q_s$ is the inverse size of the dipole for which absorptive corrections become important.) This indicates that parton-parton correlations are too strong in the present model. Therefore we consider our results are close to the maximum possible gap suppression.

\section{Possible role of NLO effects}

\subsection{Threshold factors}

The evolution equation for $\Omega^a_k$, (\ref{eq:evol1}), and the analogous one for $\Omega^a_i$, are written in the leading ln$(1/x)$ approximation, without any rapidity threshold. The emitted parton, and correspondingly the next rescattering, is allowed to occur just after the previous step. On the other hand, it is known that a pure kinematical $t_{\rm min}$ effect suppresses the probability to produce two partons close to each other. Moreover, this $t_{\rm min}$ effect becomes especially important near the production vertex of the heavy object. It is therefore reasonable to introduce some threshold rapidity gap, $\Delta y$ and to compute $S^2_{\rm enh}$ only allowing for absorption outside this threshold interval. The results of Fig.~\ref{fig:SSb} were calculated for $\Delta y=1.5$. The same rapidity interval was used to separate low- and high-mass diffractive dissociation in the model \cite{KMRnns}. Below, we present the results for the survival factors for $\Delta y =0,~1.5,~2.3$. For $\Delta y=2.3$ all the NLL BFKL
corrections \cite{BFKLnnl} may be reproduced by the threshold
effect \cite{thr,JHEP,kkmr,salam}.  Technically we can allow for a non-zero $\Delta y$ by just shifting $y_j$ to $y_j={\rm ln}1/\xi_j~-\Delta y$.

For the CED production of a high $E_T$ photon pair at the Tevatron with the ``threshold'' $\Delta y=2.3$ ($\Delta y=0$) we reproduce only one half (1/7) of the value of the full gap survival factor $S^2_{\rm eff}=0.05$
that was used in \cite{KMRSgg} to predict the corresponding cross section. For lighter CED $\chi_c$ production, where the `enhanced' absorption is stronger (that is $S_{\rm enh}$ is smaller)  with $\Delta y=2.3$ we now obtain\footnote{We assume that the non-perturbative contribution is caused by the largest-size Pomeron component $P_1$, while the perturbative contribution is due to the second component $P_2$.} about factor 0.35 smaller prediction than that published in \cite{KMRSchi}.
For the CED Higgs boson production at the LHC, the model gives $S^2_{\rm eff}=0.004,\; 0.09$ and 0.015 for $\Delta y=0,\; 1.5$ and 2.3 respectively.

The existing data from the Tevatron for exclusive $\gamma \gamma$ \cite{CDFgg}, dijet \cite{CDFdijet} and $\chi_c$ \cite{CDFchi} production are in agreement with the predictions \cite{KMRSgg,KMRSchi}, indicating that actually the value of $\langle S^2_{\rm enh} \rangle$ is larger than obtained by the `soft' model of \cite{KMRnns}. So, we should regard
\be
\langle S^2_{\rm eff} \rangle~=~0.015 \pm 0.01
\ee
as a conservative (lower) limit for the gap survival probability in the exclusive production of a Higgs boson of mass 120 GeV at the LHC.  Recall that this effective value should be compared with $S^2$ obtained using the exponential slope $B=4~{\rm GeV}^{-2}$.

\subsection{Black Disc Regime (BDR)}

Note that, at relatively low scales, $Q^2 \lapproxeq 3~{\rm GeV}^2$, the NLO gluon density extracted from the global parton analyses \cite{global} does not increase with $1/x$. Since the absorptive cross section $\sigma_{\rm abs} \sim \alpha_s /Q^2$ decreases with increasing $Q^2$ faster than the gluon density grows ($xg \sim (Q^2)^\gamma$ with $\gamma<0.5$), the role of larger scales is even smaller. Given this phenomenological observation, we may assume that, for $\sqrt{s}>1$ TeV, the parton density  which controls the `enhanced' absorptive effect, does not grow with increasing energy. Hence the $S^2_{\rm enh}$ factor will depend {\it only} on the available rapidity interval for enhanced rescattering; that is, on the $M_A^2/s$ ratio. but {\it not} on the initial energy. 

This observation clearly contradicts the assumption that the black disc regime (BDR) will have been reached at the LHC energy \cite{BDR}; and that the low $x$ gluon density will be so large that only on the far periphery of the proton, $b>1$ fm, will there be a chance to avoid an additional inelastic interaction, so that the gap survives.

Recall that the BDR of \cite{BDR} was obtained using {\it leading order} (LO) gluons which grow steeply with $1/x$. This growth is simply an artefact of the absence of a LO coefficient function, $C_{\gamma g}^{(0)}=0$, corresponding to $\gamma g$ fusion in deep inelastic scattering (DIS), and that there is {\it no} $1/z$ singularity in the LO quark-quark splitting function, $P_{qq}(z)$. In order to reproduce the DIS data, these deficiencies are compensated by an artifically large gluon density at low $x$ values\footnote{We thank Robert Thorne for discussions.}. When the NLO (and NNLO) contributions are included the DIS data are described by a flat (or even decreasing  with $1/x$) gluon at the relevant moderately low scales. With such a flat distribution, we expect that the value of $S^2_{\rm enh}$ for central exclusive Higgs production at the LHC will be {\it larger} than that for the exclusive production of a pair of high $E_T$ photons at the Tevatron
\be
S^{\rm LHC}_{\rm enh}(M_H>120 ~{\rm GeV})~~>~~S^{\rm Tevatron}_{\rm enh}(\gamma\gamma;E_T>5~\rm{GeV)~~>~~S^{\rm Tevatron}_{\rm enh}(\chi_c)}.
\ee

Actually, in the preceding model of soft interactions \cite{KMRnns}, the parton density grows slowly with energy. After the absorptive factors $e^{-\lambda\Omega/2}$ are included in the evolution (\ref{eq:evol1}), the power growth of the opacity $\Omega \sim s^\Delta$ is replaced by the logarithmic asymptotic behaviour $\Omega \sim {\rm ln~ln}s$ \cite{KMRns}. 
As a result, instead of a strong BDR suppression, the value of $\langle S^2_{\rm enh} \rangle$ decreases by only 20-30\% when going from the Tevatron to the LHC energy, keeping the ratio $M_A/\sqrt{s}=0.01$ fixed. 

In our model we do not specify the nature of the soft partons. Howver, when we introduce BFKL we imply the partons are gluons, and so the screening exponents were chosen assuming gluons in both Pomerons.
In terms of QCD, at small $x$ and low scales, the increase of the parton density with decreasing $x$ found in our model may reflect the growth of the NLO sea quark density\footnote{At NLO the gluon density in this domain is approximately flat in $x$.} with decreasing $x$. This growth is less steep than that of the LO gluons, but is not negligible.

Note that the central exclusive amplitudes of interest are driven by gluon-gluon fusion: for example, $gg \to H,~gg \to \gamma\gamma$ and $gg \to \chi_c$. If, as noted above, there is a screening (NLO) sea-quark contribution in these amplitudes, then the effective colour factor is smaller. 

So, actually, in Nature, we expect a {\it larger} value of $S_{\rm enh}$.  First, because of the colour-factor suppression when the gluons are screened by quarks, which was not included in \cite{KMRnns}. Secondly, due to a much stronger `hot-spot' parton-parton correlation in $b$-space in the model of \cite{KMRnns} than that indicated by the HERA data.

\section{Comparison with exclusive Tevatron data}

This conclusion is confirmed by the CDF exclusive data obtained at the Tevatron. The recent observations of central exclusive $\gamma\gamma$ \cite{CDFgg}, dijet \cite{CDFdijet} and $\chi_c$ \cite{CDFchi} production are in overall agreement with the old KMR predictions based on the global soft model of 2000 \cite{oldsoft}. In detail, the data exceed the predictions \cite{KMRSgg} for $\gamma\gamma$ production, although here the present event rate is very low; whereas the predictions \cite{KMRSchi}, based on the KMR2000 model, are a bit above the observed $\chi_c$ production. 

How does the present soft model \cite{KMRnns}, which includes enhanced rescattering effects change the predictions? Clearly, the overall gap survival probability $S^2$ will be smaller than that based on KMR2000. However, the difference is not large for the exclusive production, $pp \to p+A+p$, of a heavy system $A$. To be precise, for $M_A=120$ GeV, the `old' soft survival prefactors are\footnote{The results, based on KMR2000 \cite{oldsoft}, in the form $\langle S^2 \rangle \langle p^2_t \rangle^2$ were published in \cite{KMRSchi}.}
$$\langle S^2 \rangle \langle p^2_t \rangle^2 = 0.0015  {\rm ~~at~ the~   LHC ~~~~(KMR 2000)}$$
$$\langle S^2 \rangle \langle p^2_t \rangle^2 = 0.0030  {\rm ~~at~ the~   Tevatron,~~~~~~~~~~}$$
while the present soft model \cite{KMRnns}, which includes enhanced recattering, gives
$$\langle S^2 \rangle \langle p^2_t \rangle^2 = 0.0010  {\rm ~~at~ the~   LHC ~~~~(KMR 2008)}$$
$$\langle S^2 \rangle \langle p^2_t \rangle^2 = 0.0025  {\rm ~~at~ the~   Tevatron,~~~~~~~~~~}$$
where all these numbers have units of $({\rm GeV})^4$.
We see that the inclusion of enhanced rescattering does reduce the exclusive $pp \to p+A+p$ cross section, but well within the uncertainty. However, for the relatively light $\chi_c$ the available rapidity interval for the enhanced suppression is large. In this case we find that the inclusion of enhanced rescattering causes a larger difference -- reducing the exclusive cross section by about $\frac{1}{3}$. As a result, the new `prediction', which includes $S_{\rm enh}$, is a bit below the CDF data on central exclusive $\chi_c$ production \cite{CDFchi}, rather than a bit above as for our old prediction \cite{KMRSchi} based on KMR2000 \cite{oldsoft}. 

Recall that there is no theoretical or phenomenological reason to have a strong energy dependence of the gap survival factor $S^2$. So the CDF central exclusive data measured at the Tevatron can be used to check and to confirm the central exclusive cross sections predicted at the LHC.
Forthcoming higher-statistics data from the Tevatron will be valuable to further check our predictions for exclusive processes at the LHC.

\section{Summary}

We briefly list the main findings of our analysis.
\begin{itemize}
\item Enhanced rescattering of intermediate partons occuring in an exclusive amplitude is not negligible. It violates soft-hard factorization, modifying the input parton distributions used in the QCD calculation of the bare $pp \to p+A+p$ amplitude, in comparison with those measured in deep inelastic scattering.
\item However, the suppression caused by $S_{\rm enh}^2$ is numerically not large for the exclusive production of a heavy mass system (like a Higgs boson) at the LHC.
\item We noted two results of NLO analyses of HERA data. First, `flat' small-$x$ gluons are obtained at the relevant scales in the global fits of deep inelastic and related data. Second, small values of the saturation scale $Q_s(b)$ are obtained for impact parameters larger than about 0.6 fm. These observations imply that the gluon density, which gives rise to the enhanced absorptive correction, is not large, and that the black disc regime is not reached at the LHC.
\item The energy dependence of the gap survival factor is smooth. Therefore exclusive $p{\bar p} \to p+A+{\bar p}$ data observed at the Tevatron can be used to evaluate the suppression expected at the LHC.
\item The present Tevatron data imply that the predictions given here (and in related papers) should be considered as conservative lower limits. This offers the possibility for the observation of Higgs bosons in exclusive processes at the LHC.
\end{itemize}

\section*{Acknowledgements}

We thank Mike Albrow, Aliosha Kaidalov, Risto Orava, Andy Pilkington and Robert Thorne for useful discussions.
MGR thanks the IPPP at the University of Durham for hospitality.
The work was supported by  grant RFBR
07-02-00023, by the Russian State grant RSGSS-3628.2008.2.

\end{document}